\documentclass[preprint]{aastex}

\slugcomment{Accepted by the ApJ}

\shorttitle{TEMPERATURE EVOLUTION}

\shortauthors{Song et al.}

\begin{document}
\title{TEMPERATURE EVOLUTION OF MAGNETIC FLUX ROPE IN A FAILED SOLAR ERUPTION }
\author{H. Q. SONG\altaffilmark{1,2}, J. ZHANG\altaffilmark{2,3,4}, X. CHENG\altaffilmark{3},  Y. CHEN\altaffilmark{1}, R. LIU\altaffilmark{4}, Y. M. WANG\altaffilmark{4}, AND B. Li\altaffilmark{1}}

\affil{1 Shandong Provincial Key Laboratory of Optical Astronomy
and Solar-Terrestrial Environment and Institute of Space Sciences,
Shandong University, Weihai, Shandong 264209, China}

\affil{2 School of Physics, Astronomy and Computational Sciences,
George Mason University, Fairfax, Virginia 22030, USA}
\email{jzhang7@gmu.edu}

\affil{3 School of Astronomy and Space Science, Nanjing
University, Nanjing, Jiangsu 210093, China}

\affil{4 CAS Key Laboratory of Geospace Environment, Department of
Geophysics and Planetary Sciences, University of Science and
Technology of China, Hefei, Anhui 230026, China}

\begin{abstract}
In this paper, we report for the first time the detailed
temperature evolution process of the magnetic flux rope in a
failed solar eruption. Occurred on January 05, 2013, the flux rope
was impulsively accelerated to a speed of $\sim$400 km/s in the
first minute, then decelerated and came to a complete stop in two
minutes. The failed eruption resulted in a large-size high-lying
($\sim$100 Mm above the surface) high-temperature ``fire ball"
sitting in the corona for more than two hours. The time evolution
of the thermal structure of the flux rope was revealed through the
differential emission measure analysis technique, which produced
temperature maps using observations of the Atmospheric Imaging
Assembly on board \textit{Solar Dynamic Observatory}. The average
temperature of the flux rope steadily increased from $\sim$5 MK to
$\sim$10 MK during the first nine minutes of the evolution, which
was much longer than the rise time (about three minutes) of the
associated soft X-ray flare. We suggest that the flux rope be
heated by the energy release of the continuing magnetic
reconnection, different from the heating of the low-lying flare
loops, which is mainly produced by the chromospheric plasma
evaporation. The loop arcade overlying the flux rope was pushed up
by $\sim$10 Mm during the attempted eruption. The pattern of the
velocity variation of the loop arcade strongly suggests that the
failure of the eruption be caused by the strapping effect of the
overlying loop arcade.
\end{abstract}


\section{INTRODUCTION}
Magnetic flux ropes play an important role in solar eruptions
manifested as coronal mass ejections (CMEs) and/or flares.
Recently, a new line of observational structures, namely EUV
hot-blobs and/or channels, have been proposed to be the most
direct manifestation of flux ropes (Cheng et al. 2011; Zhang et
al. 2012; Cheng et al. 2013; Patsourakos et al. 2013) with the
Atmospheric Image Assembly (AIA; Lemen et al. 2012) on board the
\textit{Solar Dynamics Observatory (SDO)} (Pesnell et al. 2012).
In particular, Zhang et al. (2012) revealed the flux rope as a
conspicuous hot channel structure prior to and during a solar
eruption, which initially appeared as a writhed sigmoid with a
temperature as high as $\sim$10 MK, then continuously transformed
itself toward a semi-circular shape and acted as the essential
driver of the resulting CME.

While previous studies have been focusing on the morphological and
kinematic evolution of magnetic flux ropes, little is known about
their  thermal evolution. The scenario of heating solar flare
loops is well accepted in classical flare models (Carmichael 1964;
Sturrock 1966; Hirayama 1974; Kopp \& Pneuman 1976), in which
flare loops are mainly heated by the thermal plasma from the
chromospheric evaporation (Doschek et al. 1980; Feldman et al.
1980). Now the question is whether flux ropes, seen as expanding
hot channels in a much higher corona, are heated by the same
process. In this letter, we intend to use
differential-emission-measure (DEM) based temperature analysis
method on an event of failed flux rope eruption to address this
issue. Recently, DEM analyses have been applied to diagnose the
physical properties of CMEs (Zhukov \& Auch\`{e}re 2004; Landi et
al. 2010; Tian et al. 2012; Cheng et al. 2012; Tripathi et al.
2013), the quiet Sun (V\'{a}squez et al. 2010) and the post-flare
loop systems (Reeves \& Weber 2009; Warren et al. 2013). The
multi-passband broad-temperature capability of AIA makes it ideal
for constructing DEM models. We find that this method is
particularly useful in studying the failed solar eruption, in
which the thermal structure is conspicuous and long lasting in AIA
field of view (FOV).

There have been several studies on failed eruptions, but the
physical cause of the failure remains elusive. It is found that
long duration flares tend to be more eruptive (Kahler et al.
1989), but many  other factors could be invovled, such as the
interaction between filaments and their associated magnetic
environment (Ji et al. 2003; Williams et al. 2005; Gibson \& Fan
2006; Jiang et al. 2009; Kuridze et al. 2013), the degree of the
helical twist in the filament (Rust \& LaBonte 2005), the
overlying arcade field (Wang \& Zhang 2007), and the position of
the reconnection site (Gilbert et al. 2007). Numerical simulations
showed that the kink instability could trigger a failed filament
eruption if the overlying magnetic field decreases slowly with
height (e.g., T\" or\" ok \& Kliem, 2005).

For the failed eruption in this paper, we can clearly observe and
track the flux rope structure, as a hot EUV blob, before, during
and after the eruption. Using advanced DEM-based temperature map
method, we reveal, for the first time, the detailed thermal
evolution of the flux rope, and conclude that the thermal
structure is caused by the direct heating from the energy release
of the magnetic reconnection in the flux rope. Further, we find
convincing evidences that the failure of the eruption is caused by
the strapping effect of the overlying magnetic loop arcade. In
Section 2, we present the observations and results, which are
followed by a summary and discussions in Section 3.

\section{OBSERVATIONS AND RESULTS}

\subsection{Instrument and Method}

The AIA images the multi-layered solar atmosphere through 10
narrow UV and EUV passbands almost simultaneously with high
cadence (12 seconds), high spatial resolution (1.2 arcseconds) and
large FOV (1.3 R$_\odot$). The temperature response functions of
their passbands indicate an effective temperature coverage from
0.6 to 20 MK (O'Dwyer et al. 2010; Del Zanna et al. 2011; Lemen et
al. 2012;). During eruptions, the 131~\AA\ and 94~\AA\ passbands
are more sensitive to the hot plasma from flux ropes and flare
loops, while the other passbands are better at viewing the cooler
leading front and dimming regions (e.g., Cheng et al. 2011; Zhang
et al. 2012).

The observed flux $F_{i}$ for each passband can be determined by

\begin{displaymath}
F_{i}=\int R_{i}(T)\times DEM(T)dT
\end{displaymath}

\noindent where the $R_{i}(T)$ and $DEM(T)$ are the temperature
response function of passband $i$ and the plasma DEM in the
corona, respectively. Cheng et al (2012) and we use the
``xrt\_dem\_iterative2.pro" routine in SSW package to compute the
DEM. This code was originally designed for \textit{Hinode}$/$X-ray
Telescope data (Golub et al. 2004; Weber et al. 2004) and was
modified slightly to work with AIA data (Schmelz et al. 2010,
2011a, 2011b; Winebarger et al. 2011; Cheng et al. 2012). More
details and tests of this method were discussed in the Appendix of
Cheng et al. (2012). Here, we  use the DEM-weighted average
temperature per pixel defined in the following formula (Cheng et
al. 2012) to construct the temperature map in spatial domain and
study their temperature evolution in time.

\begin{displaymath}
\overline{T}=\frac{\int DEM(T)\times TdT}{\int DEM(T)dT}
\end{displaymath}

\noindent

Errors in DEM inversion arise from the uncertainties in the
response function $R_{i}(T)$ (Judge 2010) and the background
determination (e.g., Aschwanden \& Boerner 2011). According to
Cheng et al. (2012), the error of DEM-weighted temperature at the
flux rope center could be $\sim$15\%. We should point out that the
plasma, integrated along the line of sight, contains multiple
temperatures. The ``weighted" temperature we are obtaining here,
is an indicator of the overall thermal trend of the plasma in the
temperature range the instrument is sensitive to.

\subsection{The Event and Its Temperature Evolution}

On 2013 January 5, an M1.7 class soft X-ray flare occurred at the
northeast limb of the Sun, which started at 09:28 UT and peaked at
09:31 UT. The flare was located at $\sim$N20E88 (NOAA 11652) from
the perspective of the Earth. No associated CME was observed by
the Large Angle Spectroscopic Coronagraph (LASCO; Brueckner et al.
1995) and the Sun Earth Connection Coronal and Heliospheric
Investigation (SECCHI; Howard et al. 2008). But in AIA~131~\AA\
and 94~\AA\ passbands, an obvious structure with high temperature
was observed to erupt, but stopped to rise further. In Figure 1,
we present the observations of this failed eruption process.
Panels (a)-(c) are observations from 131~\AA\ , (d) is a composite
image of 94~\AA\ (Blue channel) and 171~\AA\ (Green channel) and
(e)-(f) are composite images of 131~\AA\ (Red channel) and
171~\AA\ . (See supplementary Movies 1 and 2 for the entire and
continuous eruption process in six AIA EUV passbands). The left
panels show the coronal images immediately before the eruption.
The upper-pointing red arrows indicate the position of the flux
rope structure that was about to rise and erupt. The flux rope is
better seen in 94~\AA\ in the early time of the evolution, which
means that its initial temperature should be around 6 MK; this is
consistent with the DEM analysis result shown in Figure 2 (a).
This temperature effect explains why the flux rope is better seen
in 94~\AA\, but less visible in 131~\AA\ (Figure 1 (a)) and
completely invisible in 171~\AA\ (Figure 1(d)) in the early time.
This flux rope can be observed for its favorable orientation. It
seems that the axis of the rope lied along the line of sight at
the limb of the Sun, making it a bright blob-like structure. At
09:28:56, a hot and fast narrow jet started to appear in 131~\AA\
images, and became obvious at 09:29:20 as depicted with blue
arrows in Figures 1(b) and (e). With DEM analysis, we learned that
the hot jet temperature is around 8 MK (Figures 2(b) and (c)).
Therefore, the jet is better seen in 131~\AA\, but could not be
observed in 171~\AA\ passband, see Figure 1(e).

The flux rope started to rise at about 09:29:20 UT, stopped at
09:32:35 UT, and then stayed in the position for several hours
before faded away. The flux rope could not be observed in 171~\AA\
during the entire process because of its high temperature. But the
171~\AA\ observations show the movement of the overlying loop
arcade clearly (bottom panels in Figure 1). The asterisks in the
panels show the position of the inner edge of the overlying loop
arcade. With a horizontal white line passing through the two
asterisks in the frames of earlier times when the loop arcade had
not begun to rise, it is easy to notice that the loop arcade was
shifted and raised to a higher position following the eruption.
The 211~\AA\ and 193~\AA\ observations also show the overlying
loops stressed by the rope (See supplementary Movie 1). The red
plus symbols shown in the right panels point to the apex position
of the flux rope.

Figure 2 shows a sequence of temperature maps throughout the
entire thermal evolution process of the event. Before the
eruption, the temperature of the flux rope (depicted with a white
arrow) is over 5 MK as shown in Figure 2(a). Figures 2(b) and (c)
show the hot jet (depicted with the white arrows) underneath the
flux rope. The appearance of the hot jet seemed to signal the
onset of the flare and also the onset of the flux rope rising. We
find that the temperature over the eruption region increased
quickly (See supplementary movie 3). Two sub-regions, as depicted
with a large square in the high corona and a small rectangle close
to the solar surface in Figures 2(b), (e) and (h), are selected to
be the regions of the flux rope and flare loops respectively, for
the purpose of tracking their temperature evolutions. To obtain
the characteristic average temperature of the entire flux rope,
all the pixels with temperatures greater than 5.0 MK in the
selected square are regarded as the flux rope pixels, and their
average temperature is regarded as the characteristic temperature.
The same way is used to get the average temperature of the flare
loop region as selected with the rectangle.

The temperature evolution of the flux rope and the flare loops are
shown in Figure 3 with red and blue solid line, respectively,
along with the \textit{Geostationary Operational Environmental
Satellite (GOES)} soft X-ray 1-8~\AA\ profile in the black line.
The insert panel in Figure 3 is an enlarged portion from 09:28:08
to 09:39:08 UT to better show the relations at the beginning of
the eruption. Apparently, the temperature profile of the flare
loops is almost the same as the soft X-ray profile, i.e., the same
peak time and the similar rise phase. This is consistent with the
classic thick-target model that the flare region is heated by the
precipitation of energetic electrons accelerated in flare
reconnection; Note that we might under-estimate the temperature of
the flare loops for the following two reasons: first, a part of
the hot near-footpoint sources might be occulted by the limb;
second, a small fraction of the flare region were saturated in AIA
images. Nevertheless, we don't believe that these effects will
change the temporal profile of the temperature evolution. However,
the temperature evolution of the flux rope is remarkably different
from that of the flare loops, i.e., the duration of the
temperature increase lasted much longer than that in the flare
loops and the rise phase of the soft X-ray flare. For clarity of
discussion, we may divide this duration into two phases:
impulsive-heating phase and post-impulsive-heating phase. In the
impulsive-heating phase, the temperature increased from
$\sim$5.2MK ($\sim$09:29 UT) to $\sim$8.5 MK ($\sim$09:31 UT),
while the temperature increased from $\sim$8.5 MK to $\sim$10 MK
($\sim$09:38 UT) during the post-impulsive-heating phase. Then the
temperature began to slowly decrease to $\sim$5.6 MK until 12:00
UT. It is obvious that the impulsive and post-impulsive heating
phases correspond well with the rise phase and decay phase of the
associated flare.

The continuing rise of temperature in the flux rope after the
flare rise phase is an interesting discovery in this study. It
seems that the heating of the flux rope should be different from
that of the flare loops. We suggest that the direct thermal
heating from the magnetic reconnection region, without via the
chromospheric evaporation, should be responsible for the heating
of the flux rope. The continuing strong heating of the flux rope
after the flare soft X-ray peak time is related with the
continuing energy release during the decay phase of the soft X-ray
flare. In this long duration flare, there should be continuing
magnetic reconnection in the corona. Further, the failure of the
eruption resulted in less energy converted to the kinetic energy
of the bulk plasma from the released magnetic energy; this extra
energy should further heat the flux rope trapped in the corona.
Consequently, the eruption produced a conspicuous large-size
high-lying ($\sim$100 Mm above the surface) high-temperature
($\sim$6--10 Mk) ``fire ball" sitting in the corona for more than
two hours. In the end, the hot structure faded away and became
indistinguishable from the ambient as shown in Figure 2(i).

To further elucidate the relation between the flux rope heating
and the magnetic reconnection, we investigated the location of the
coronal hard X-ray sources using data from \textit{Reuven Ramaty
High Energy Solar Spectroscopic Imager (RHESSI)} (Lin et al 2002).
Unfortunately, RHESSI was in night until $\sim$09:46 UT. RHESSI
observations around 09:47 UT are thus selected and shown as black
contours in the 12--20 keV band at 50\%, 70\% and 90\% of the
maximum in Figures 1(c) and 2(f). The contours apparently show a
double X-ray source. It has been suggested that the region of
magnetic reconnection should be between the two X-ray sources
(e.g., Liu et al. 2008).  Therefore, it seems that the magnetic
reconnection separates the flux rope structure into two parts: the
upper part and the lower part. Such scenario of  magnetic
reconnection leading to partial eruption have been reported in
several studies (Gilbert et al. 2000; Gibson \& Fan 2006; Tripathi
et al. 2007; Sterling et al. 2011; Tripathi et al. 2013).

\subsection{Kinematic Evolution of The Failed Eruption}

We also study the kinematic evolution process of the flux rope and
its overlying loops. The heights of flux rope apexes (red plus
symbols shown in the right panels of Figure 1) and the inner edge
of the overlying loop arcades (blue asterisk symbols shown in the
bottom panels of Figure 1) are tracked in the AIA images and shown
in Figure 4(a). Their velocities are shown in Figure 4(b), along
with the \textit{GOES} soft X-ray 1-8 \AA\ profile shown in the
black line. The velocities are calculated from the numerical
differentiation using the 3-point Lagrangian interpolation of the
height-time data. Note that the uncertainties of the velocities
come mainly from the uncertainties in the height measurement. The
measurement errors are estimated to be 2 pixels. These errors are
propagated in the standard way to estimate the errors of velocity.
The red, blue and black filled circles show the times of the
maximum velocities of the flux rope and overlying loops, and the
peak of the soft X-ray flux, respectively.

From Figure 4(a) we find that the flux rope started to show
noticeable movement at 09:29:20 UT, and the loop arcade, started
to move at a later time at 09:30:11 UT. Apparently, the ascending
motion of the overlying loop arcade is induced by the rising
motion of the erupting flux rope; the early interaction between
the flux rope and the loop arcade should take place at a time
between 09:30:11 UT and 09:30:23 UT as indicated by the two
vertical dashed lines in Figure 4; at this time, the flux rope had
already obtained a speed of about 400 km/s. The interaction
apparently prevented the further eruption of the flux rope and led
its velocity to decrease quickly, until came to a full stop two
minutes later. Before the contact of the interaction, the velocity
of the flux rope increased from $\sim$95 km s$^{-1}$ at 09:39:20
UT to $\sim$431 km s$^{-1}$ at 09:30:08 UT; during this 48 s of
``free" acceleration, the average acceleration rate is estimated
to be $\sim$7000 m s$^{-2}$. This is an extremely strong
acceleration, when compared with the rates of most CMEs which are
typically lower than 1000 m s$^{-2}$ (Zhang \& Dere 2006). The
interaction made the velocity of the rope just increased from 431
km s$^{-1}$ at 09:30:08 UT to 438 km s$^{-1}$ at 09:30:20 UT with
an average acceleration 580 m s$^{-2}$. Then the velocities of the
flux rope began to decrease quickly from 438 km s$^{-1}$ to almost
zero at 09:32:35 UT with the average decceleration rate 3240 m
s$^{-2}$. On the other hand, the velocity of the loop arcade
started at about 70 km s$^{-1}$ when the movement became
noticeable at 09:30:11 UT, and increased to 98 km s$^{-1}$ at
09:30:23 UT as its peak value. The early acceleration of the loop
arcade was not observed, probably because the early interaction
was short and prompt, even the 12-second cadence of AIA was not
sufficient to capture the evolution. The instantaneously
accelerated loop arcade immediately began its deceleration. The
deceleration phase of the loop arcade coincided well with that of
the flux rope, coming to a full stop about two minutes later. We
conclude that the magnetic strapping effect of the overlying loop
arcade, likely due to the magnetic tension force along the loops,
made the eruption a failed one, instead of forming an escaping
CME.

\section{SUMMARY AND DISCUSSIONS}

A failed flux rope eruption associated with an M1.7 class flare
was observed by the AIA at the northeast limb of the Sun on 2013
January 5, which provided us an unprecedented opportunity for
studying the detailed temperature evolution of the flux rope. We
find that the flux rope existed in the corona with a high
temperature ($\sim$5 MK) before its eruption. The temperature
evolution of the flux rope can be divided into two phases:
impulsive-heating phase and post-impulsive-heating phase, which
correspond well with the rise phase and decay phase of the
associated flare. On the contrary, the temperature of the
low-lying flare loops only increased during the flare rise phase,
and quickly decreased during the decay phase. Furthermore, there
was a gap of low temperature between the flux rope and the flare
loops. All of these observations indicate that the flux rope
should be heated by a different process from  that of the flare
loops. The heating process of flare loops have been well
understood and widely accepted as summarized below. Energetic
electrons and ions produced by flare magnetic reconnection
precipitate from the coronal acceleration site and lose their
energy in the dense underlying chromosphere via Coulomb
collisions. The temperature in the chromosphere increases and the
resulting pressure exceeds the ambient chromospheric pressure,
which leads the heated plasma to expand along the magnetic field,
forming flare loops (Doschek et al. 1980; Feldman et al. 1980).
While for the flux rope, we suggest that it should be heated
directly by the thermal energy generated at the reconnection site
through the thermal conduction. It is unlikely that chromospheric
evaporation plays a strong role in heating the flux rope.

The velocity analysis of the flux rope and the overlying loop
arcades strongly suggests that the failure of the eruption be
caused by the strapping effect of the overlying magnetic loop
arcade. The strapping effect, or the tension force of the
line-tying field, is the determining factor that contained the
flux rope from further eruption.

\acknowledgments We thank the anonymous referee for his/her
valuable comments that helped to improve the paper. SDO is a
mission of NASA's Living With a Star Program. This work is
supported by the 973 program 2012CB825601, NNSFC grants 41104113,
41274177, 41274175, and 41331068. H. Q. Song is also supported by
the Natural Science Foundation of Shandong Province ZR2010DQ016.
J. Zhang is supported by NSF grant ATM-0748003, AGS-1156120 and
AGS-1249270. B. Li is supported by NNSFC grant 41274176.

\clearpage

\begin{figure}
\epsscale{1.0} \plotone{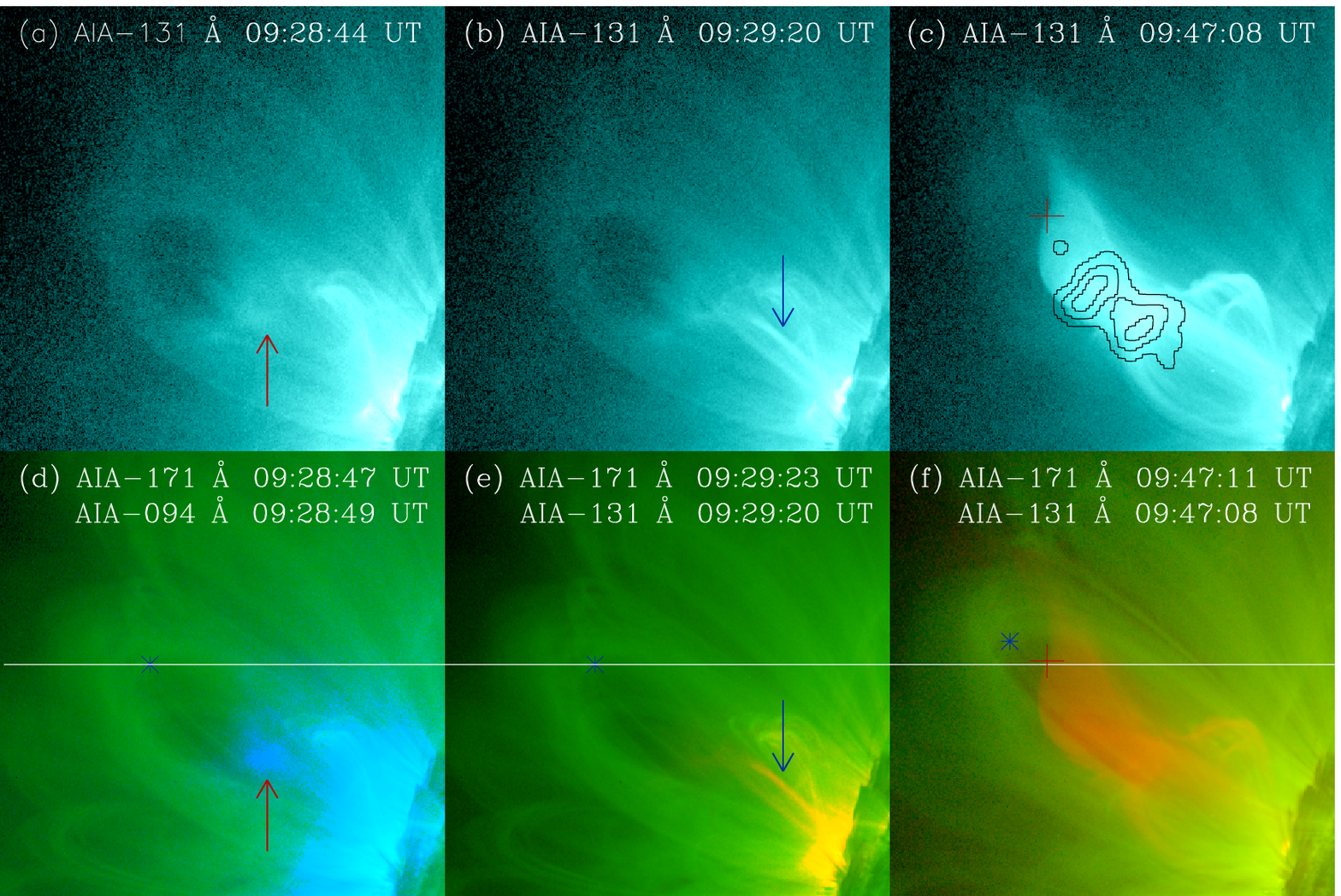} \caption{The failed flux rope
eruption on 2013 January 5. The FOV is taken to be [-1140,-890]
and [320, 570] arcsec for the horizontal and vertical axis,
respectively. (Animations of the eruption process with full AIA
cadence and a color version of this figure are available in the
online journal)\label{fig1}}
\end{figure}

\begin{figure}
\epsscale{1.0} \plotone{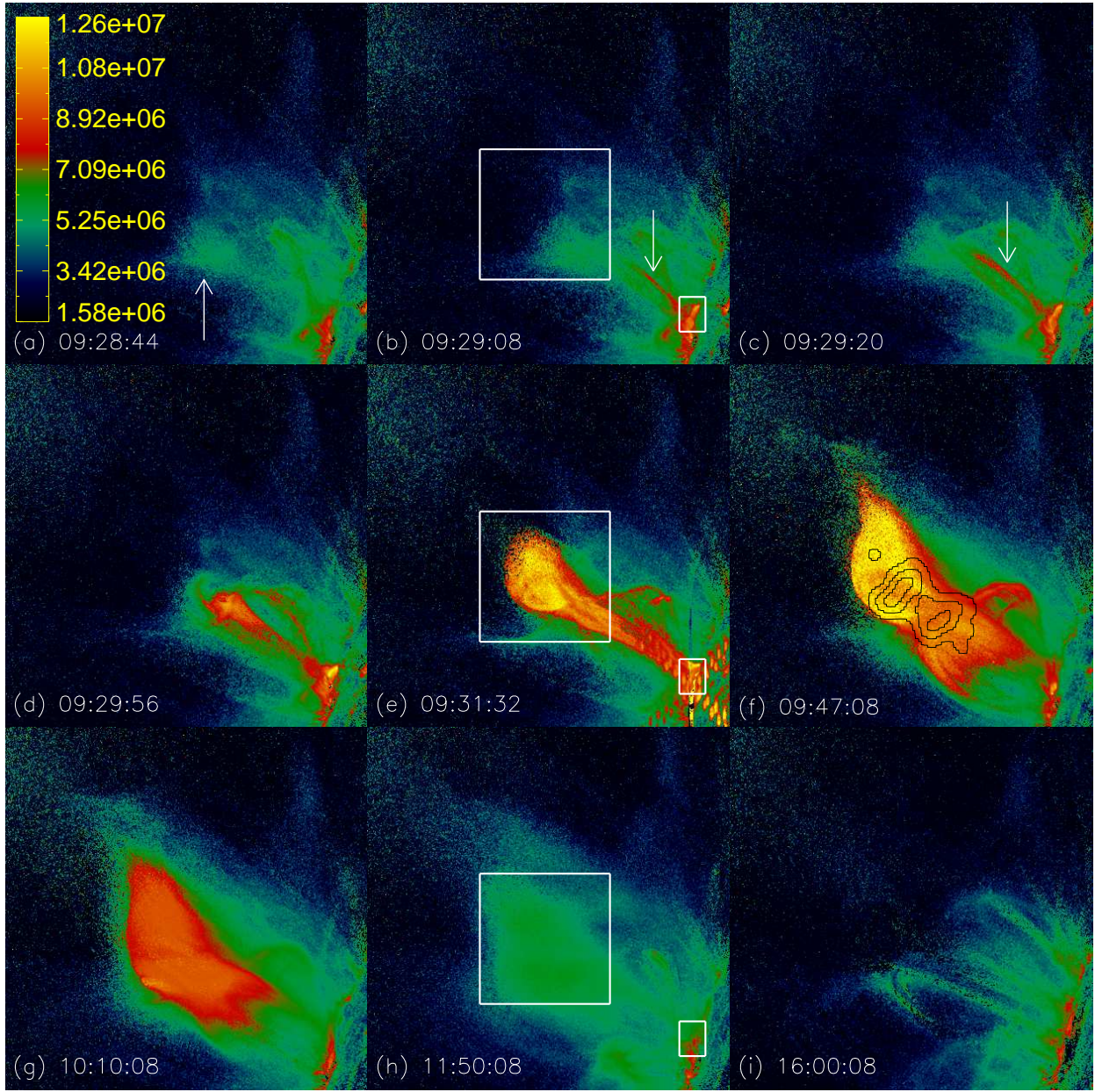} \caption{Temperature evolution
process of the failed flux rope eruption. The FOV is the same as
in Figure 1. (Animations of the full evolution of the temperature
maps and a color version of this figure are available in the
online journal.)\label{fig2}}
\end{figure}

\begin{figure}
\epsscale{1.0} \plotone{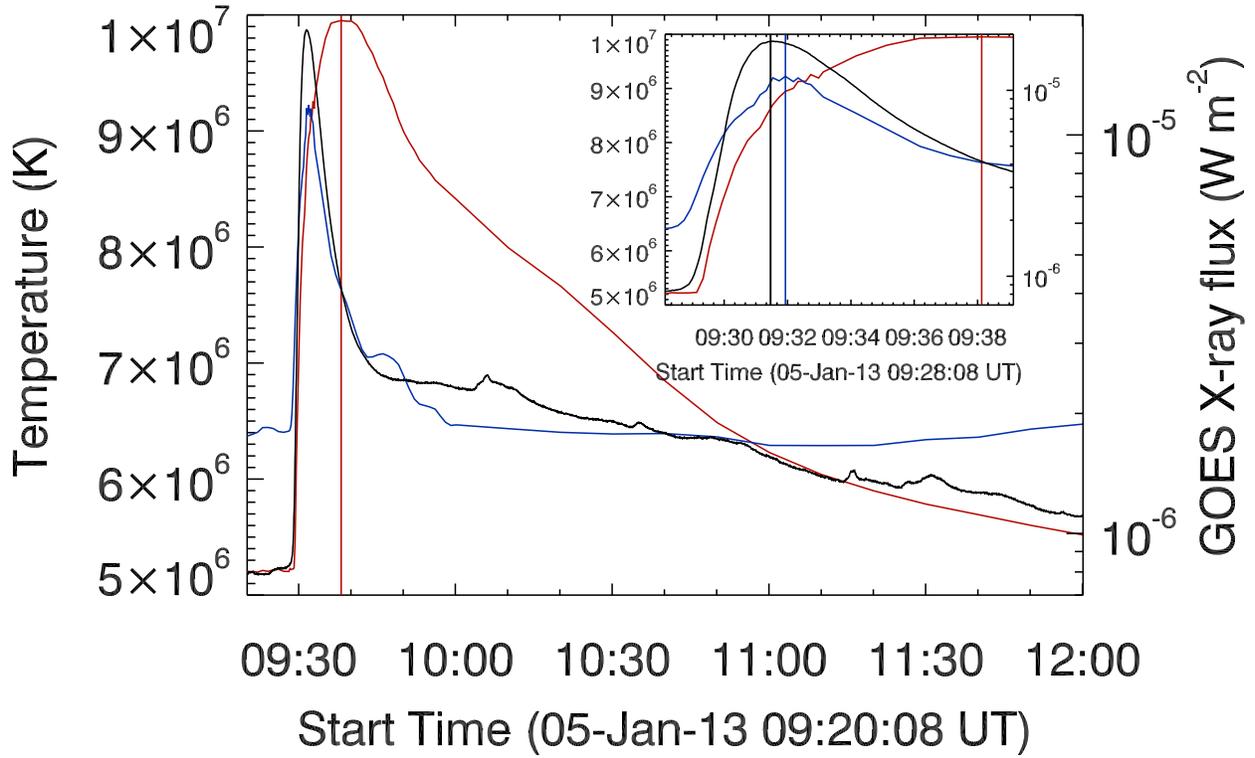} \caption{The temperature-time
profiles of the high-lying flux rope (red) and the low-lying flare
loops (blue), along with the profile of \textit{GOES} soft X-ray
1-8 \AA flux (black). (A color version of this figure is available
in the online journal.)\label{fig3}}
\end{figure}

\begin{figure}
\epsscale{1.0} \plotone{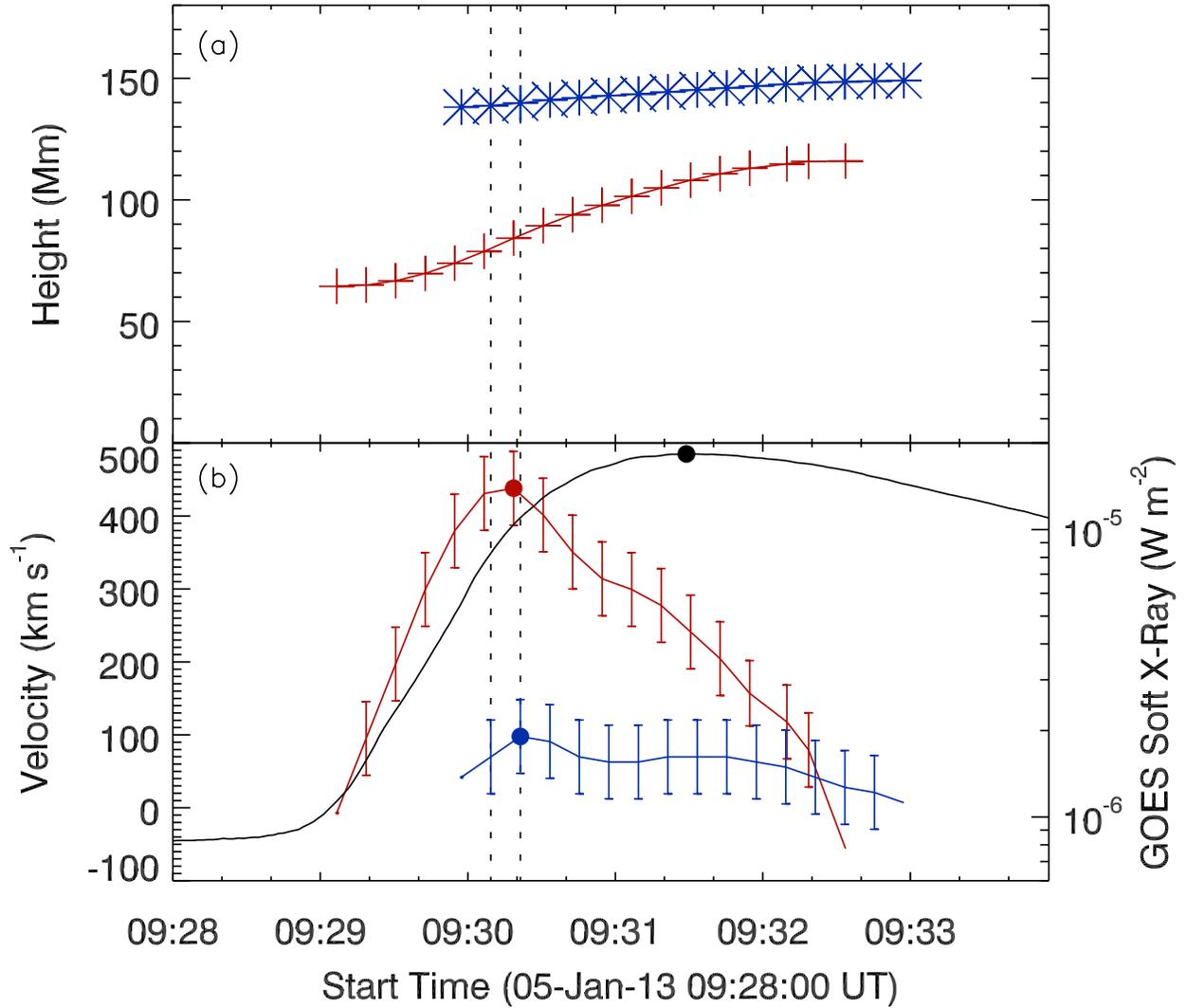} \caption{(a) The height-time
profiles of the flux rope (red) and the loop arcade (blue). (b)
The velocity-time profiles of the flux rope (red) and the loop
arcade (blue), along with the profile of \textit{GOES} soft X-ray
1-8 \AA flux. (A color version of this figure is available in the
online journal.)\label{fig4}}
\end{figure}

\end{document}